\documentclass[floats,aps,twocolumn]{revtex4-1}
\usepackage{amsmath}
\usepackage{epsfig}
\usepackage{color}
\usepackage{endnotes}
\let\footnote=\endnote

\newcommand{\be}{\begin{equation}}
\newcommand{\ee}{\end{equation}}

\newcommand{\ber}{\begin{eqnarray}}
\newcommand{\eer}{\end{eqnarray}}

\newcommand{\bs}[1]{\ensuremath{\boldsymbol{#1}}}

\begin{document}

\title{The isoscalar monopole resonance of the alpha 
       particle:\\ a prism to nuclear Hamiltonians}

\author{Sonia Bacca$^{1}$,
  Nir Barnea$^{2}$
  Winfried Leidemann$^{3,4}$, 
  Giuseppina Orlandini$^{3,4}$
  }

\affiliation{
  $^{1}$TRIUMF, 4004 Wesbrook Mall, Vancouver, B.C. V6J 2A3, Canada \\ 
  $^{2}$Racah Institute of Physics, The Hebrew University, 91904, Jerusalem, Israel\\
  $^{3}$Dipartimento di Fisica, Universit\`a di Trento, I-38123 Trento, Italy \\
  $^{4}$Istituto Nazionale di Fisica Nucleare, Gruppo Collegato di Trento,
  I-38123 Trento, Italy 
}

\begin{abstract}
We present an ab-initio study of the isoscalar monopole excitations  of $^4$He
using different realistic nuclear interactions,  
including  modern effective field theory  potentials.
In particular we concentrate  on the transition form factor $F_{\cal M}$ to the narrow $0^+$ resonance 
close to threshold. $F_{\cal M}$  exhibits  a strong potential model dependence,
and can serve as a kind of prism to distinguish among different nuclear force models.
Comparing to the measurements obtained from inelastic electron scattering off $^4$He, one finds that
 the  state-of-the-art theoretical transition form factors are at variance with experimental data, especially
in the case of effective field theory potentials.
We discuss some possible reasons for such discrepancy, which still remains a puzzle.
\end{abstract}

\bigskip


\maketitle

The isoscalar monopole strength of large nuclei has been extensively studied
since the discovery of a giant monopole resonance in $^{144}$Sm and $^{208}$Pb~\cite{YoR77}.
The reason for the great interest in such excitations originates from  their connection
to the incompressibility modulus of infinite nuclear matter~\cite{BoL79,Bl80}.
The alpha particle is a light nucleus, that however has a binding energy per particle similar to that of large systems
and a high central density. While it possesses no bound excited states, 
it exhibits a very pronounced narrow resonance ($^4$He*) with the same quantum numbers $0^+$ as the ground state,
i.e.,  an isoscalar monopole resonance. 
Today, the development of few-body theories has reached a point, where an
ab-initio calculation of the four-body isoscalar monopole transition  
strength can be carried out with high precision. 
As will become evident in the following, the comparison of such four-body results 
with experimental data can serve 
as a stringent test for nuclear Hamiltonians, that are the sole  ingredients of an  
ab-initio quantum mechanical approach.

The  four-nucleon isoscalar monopole resonance  is located  at $E_R^{exp}=-8.20\pm 0.05$ MeV, with a 
width of  270$\pm 50$ keV~\cite{Wa70},
just above the first two-body break-up threshold $E_{thr}^p=-8.48$ MeV
into a proton and a triton and  below the next 
threshold $E_{thr}^n=-7.74$ MeV into a neutron and a $^3$He. 
A summary about the experimental studies of the spectrum of  $^4$He
can be found in Ref.~\cite{FiM73}.
Valuable information about the nature of the resonance 
is given by the transition form factor $F_{\cal M}(q)$ measured in 
electron scattering experiments ($^4$He$(e,e')^4$He*) at various momentum transfer $q$. Similarly to the case of 
the elastic form factor, the $q$ dependence of $F_{\cal M}$ reflects
the dynamics at various interaction ranges.

The  progress in ab-initio few-body methods  allows today to obtain accurate results 
for observables in 
light nuclear systems using realistic potential models  (see review~\cite{LeO12}).
In recent years the  debate regarding potential models 
has boosted, especially after the introduction of the effective field theory (EFT) strategy in nuclear physics~\cite{EFT}. At present,
both phenomenological realistic and chiral EFT potentials
are used in ab-initio calculations, but only
for very few observables large differences are found, 
e.g., for the polarization observable $ A_y$ of $p\,$-$^3$He scattering~\cite{Ki11}. 
In this letter, we point out that the calculated  $^4$He isoscalar monopole resonance transition
form factor $F_{\cal M}(q)$ depends dramatically on the nuclear Hamiltonian.
Thus, it can serve as
a kind of prism to distinguish among nuclear force models.

\paragraph*{Main Results.}
The isoscalar monopole strength $S_{\cal M}(q,\omega)$ is in general a function of $q$
and the energy transfer $\omega$.
It is given by
\begin{eqnarray}
\nonumber
   &&S_{\cal M}(q,\omega)= \sum\!\!\!\!\!\!\!\int\,dn|\langle  n| {\cal M}(q)|0\rangle |^2\delta(\omega-E_n+E_0)\\
   &=& -\frac{1}{\pi}{\it Im}\langle 0| 
 {\cal M}^\dagger (q)\frac{1}{\omega-H+E_0+ i\epsilon} {\cal M} (q)|0\rangle \,,
\label{S_M}
\end{eqnarray}
where $|0\rangle ,|n\rangle $ and $E_0,E_n$ are eigenfunctions and eigenvalues of the nuclear Hamiltonian $H$, and 
\begin{equation}\label{monopole}
 {\cal M}(q)=\frac{G_E^s(q)}{2} \sum_i^A \, j_0(q r_i)\,,
\end{equation}
is the isoscalar monopole operator.
Here $G_E^s(q)=G_E^p(q)+G_E^n(q)$ is the nucleon electric isoscalar form factor~\cite{GaK71}, $\bs{r}_i$ is the nucleon's position,
and  $j_0$ is the spherical Bessel function of 0$^{th}$ order. 
The monopole strength can be written
as a sum of a resonance term $S_{\cal M}^{\rm res}$ and a non-resonant  background contribution $S_{\cal M}^{\rm bg}$,
\begin{equation}
S_{\cal M}(q,\omega) = S_{\cal M}^{\rm res}(q,\omega) + S_{\cal M}^{\rm bg}(q,\omega)\,.
\end{equation}
For a narrow resonance one defines the resonance transition form factor 
\begin{equation}
 |F_{\cal M}(q)|^2=\frac{1}{Z^2}\int d\omega  S_{\cal M}^{\rm res}(q,\omega)\,.
\end{equation}
\begin{figure}[htb]
\centerline{\resizebox*{8.5cm}{6.5cm}{\includegraphics[angle=0]{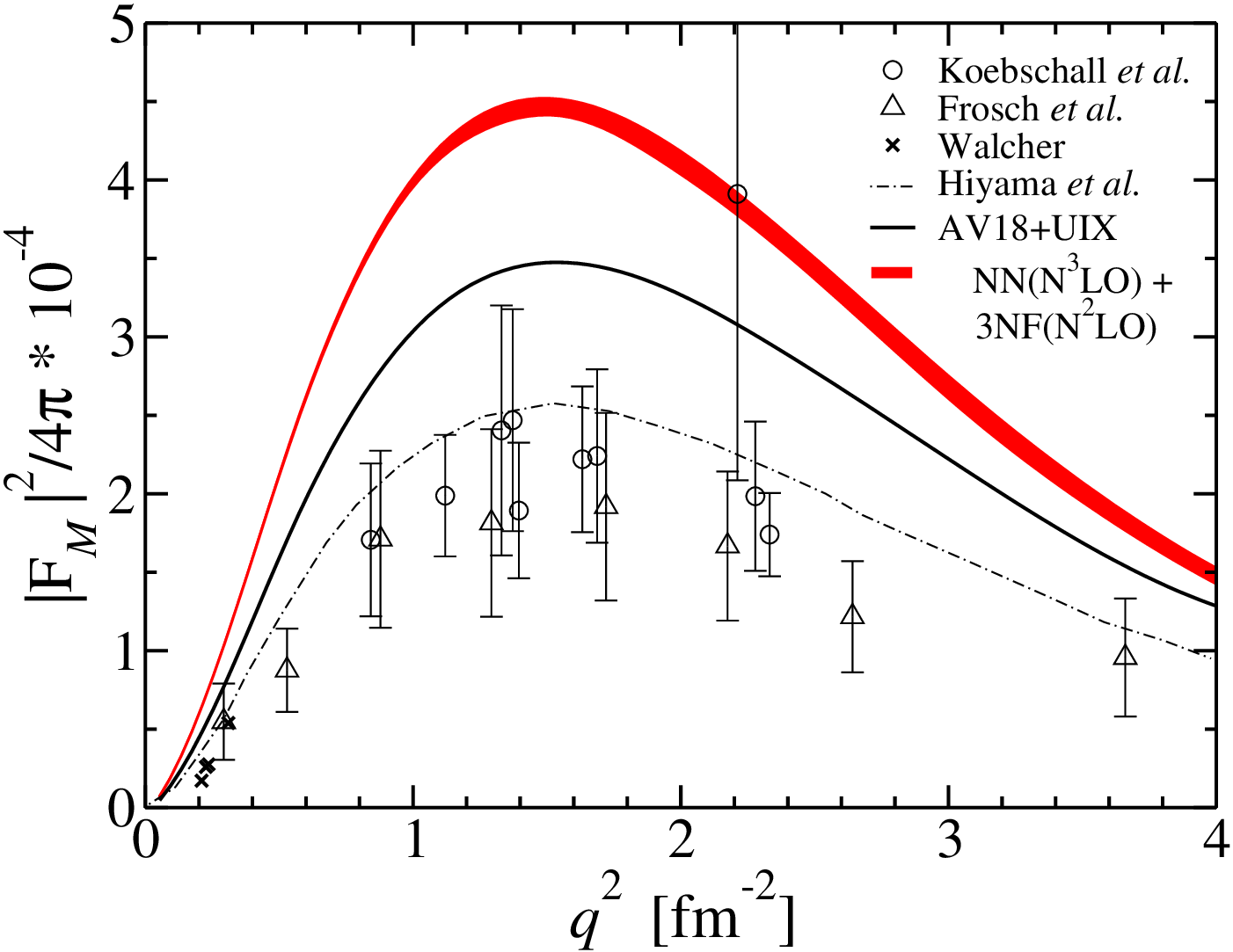}}}
\caption{(Color online) Theoretical transition form factor $|F_{\mathcal M}(q^2)|^2$ 
with $G^n_E=0$ calculated with various
force models:  AV18+UIX (full line), N$^3$LO+N$^2$LO (red band);  
result from~\cite{HiG04} (dot-dashed). Data from Frosch {\it et al.}~\cite{FrR65}, 
Walcher~\cite{Wa70} and K\"{o}bschall {\it et al.}~\cite{KoO83}.
}
\label{fig1_r}
\end{figure}

In Fig.~\ref{fig1_r}, we show results for $F_{\cal M}(q)$ with two different  Hamiltonians
including realistic three-nucleon forces (3NF)
in comparison to experimental data from inelastic electron
scattering~\cite{FrR65,Wa70,KoO83}. 
As Hamiltonians we use (i) the AV18~\cite{WiS95} NN potential plus the
UIX~\cite{PuP95} 3NF, (ii) 
an EFT based potential,
where we take the NN 
potential~\cite{EnM03}  at fourth order
 (N$^3$LO) in the chiral expansion augmented by a 3NF at order N$^2$LO~\cite{Na07}.
The Coulomb potential is taken into account in all calculations. 
Both the EFT and the AV18 NN potentials reproduce the NN scattering 
phase shifts with high precision ($\chi^2/{\rm datum}\approx 1$).
In the EFT calculations, two different parameterizations  
of the 3NF
have been used, leading to the red band in Fig.~\ref{fig1_r}. 
The chiral low energy constants $c_D$ and $c_E$ have been determined either by setting 
$c_D$ to a reasonable value and then fitting $c_E$ to the three-nucleon binding 
energies~\cite{Na07}  ($c_D=1$ and $c_E=-0.029$) or by fitting to the $^3$H binding energy 
and beta decay~\cite{GaQ09} ($c_D=-0.2$ and $c_E=-0.205$).
We also display the result of a previous calculation by Hiyama {\it et al.}~\cite{HiG04}, 
 with  the AV8' potential, a reduced version of AV18, and a simplified central 3NF, fitted to the 
binding energy of $^3$H. All three Hamiltonians reproduce the $^4$He experimental binding energy 
within one percent. 
Surprisingly, the results for $F_{\mathcal M}(q)$  strongly depend on the
Hamiltonian. Furthermore, the realistic Hamiltonians fail to reproduce the
experimental data. In particular, this is true  for the EFT forces that
predict a transition form factor twice as large as the measured 
 one. 
\begin{figure}[htb]
\centerline{\resizebox*{8.5cm}{6.5cm}{\includegraphics[angle=0]{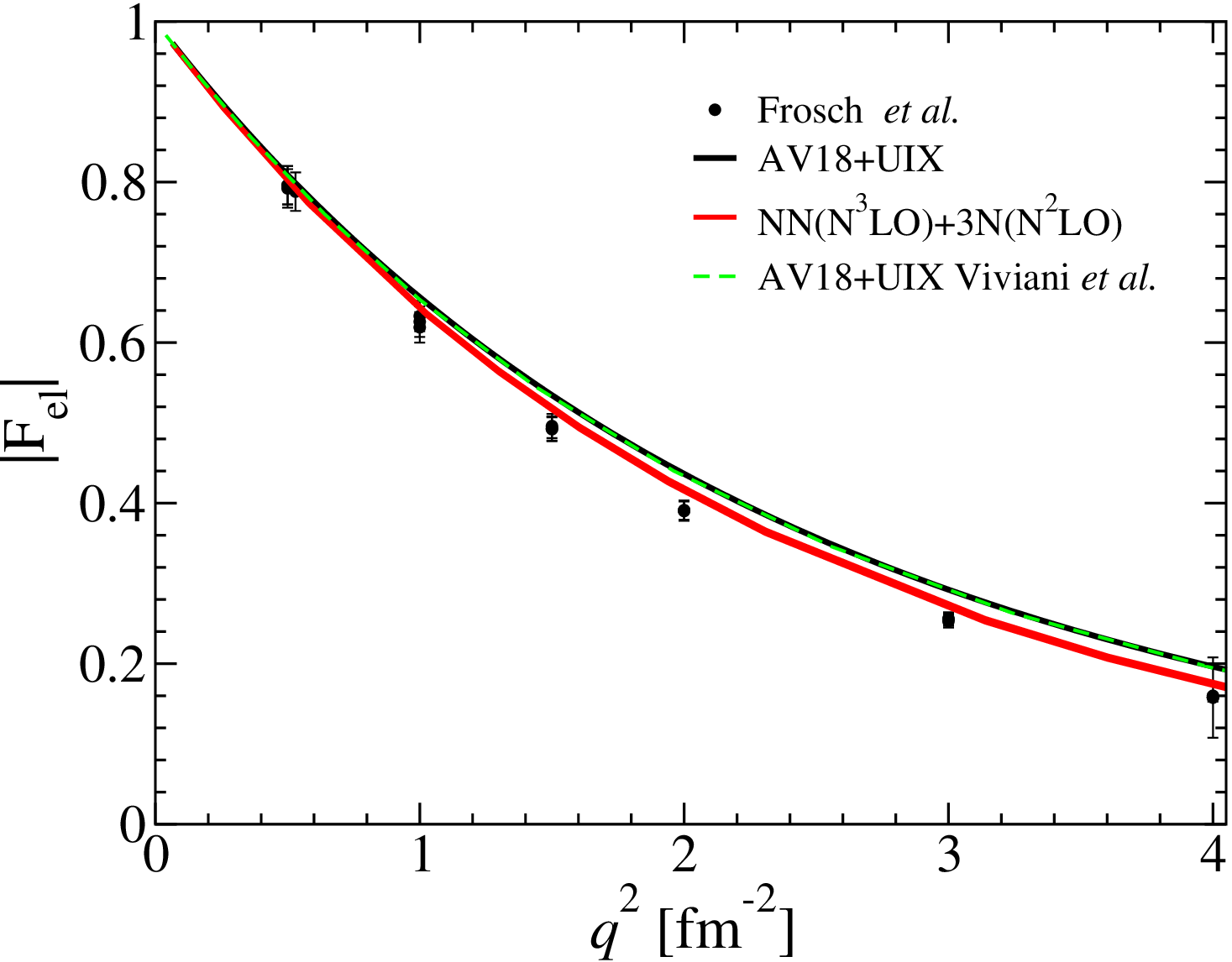}}}
\caption{(Color online) Elastic form factor $|F_{el}(q^2)|$ of $^4$He calculated with various
force models: AV18+UIX (full line); N$^3$LO+N$^2$LO (red band); result from~\cite{ViS07} with
AV18+UIX (dot-dashed). Data from Frosch {\it et al.}~\cite{FrM67}.}
\label{fig_el}
\end{figure}

In contrast, the realistic Hamiltonians lead to rather similar results 
for the elastic form factor $F_{el}(q)$  of $^4$He, defined as
\begin{equation}
F_{\rm el}(q)= \frac{1}{Z}\left<0 \left | \mathcal{M}({q}) \right| 0\right> \,.
\end{equation}
In Fig.~\ref{fig_el},
$F_{el}(q)$ is shown for the AV18+UIX model and for the chiral EFT potentials. 
The fact that the results 
do not differ significantly is not very surprising, since both Hamiltonians
give a very similar result for the radius: 1.432(2)  
fm~\cite{GaB06} for AV18+UIX and 1.464(2) fm for N$^3$LO plus the N$^2$LO of~\cite{GaQ09}  
which is not far from the experimental value
 of $1.463(6)$ fm (obtained from the charge radius of Ref.~\cite{Sic08} as explained in~\cite{correction}).
Also shown in Fig.~\ref{fig_el} is the 
 result by Viviani {\it et al.}~\cite{ViS07} with the AV18+UIX potential, which is  
indistinguishable from ours, proving the level of accuracy of contemporary four-body calculations.
\begin{table}
\caption{Ground state energies in MeV for $^3$H, $^3$He and $^4$He with N$^3$LO~\cite{EnM03} 
and N$^2$LO (parameterizations from~\cite{Na07} or~\cite{GaQ09}). Comparison of present results (EIHH) with
no core shell model (NCSM) and hyperspherical harmonics (HH) results.
\label{table_be}
}
\begin{center}
\begin{tabular}
{c c c c c}\hline\hline
 3NF from~\cite{Na07}  & EIHH & NCSM~\cite{Na07} &  HH~\cite{KiR08} & Nature \\
\hline
$^3$H  & -8.474(1)   & -8.473(5) & -8.474  & -8.48\\
$^3$He & -7.734(1)   &  & -7.733   & -7.72\\
$^4$He & -28.357(7)  & -28.34(2) & -28.37 & -28.30 \\
\hline
 3NF from~\cite{GaQ09} &  EIHH & NCSM~\cite{GaQ09} & & Nature \\
\hline
$^3$H  &  -8.472(3)  & -8.473(4)  &  & -8.48\\
$^3$He & -7.727(4)   &  -7.727(4) &  & -7.72 \\
$^4$He & -28.507(7)  & -28.50(2)  &  & -28.30\\

\hline\hline
\end{tabular}
\end{center}
\end{table}

\paragraph*{Calculational Method.}
Our calculations are based on the diagonalization of the Hamiltonian on a square integrable
hyperspherical harmonics (HH) basis.
The HH convergence is accelerated using the Suzuki-Lee unitary transformation, which
then leads to the Effective Interaction HH (EIHH) method [23,24].
The high accuracy of this approach can be inferred from the benchmark results in Ref.~\cite{KaN01}
and also here from Table I, where  we present the binding energies of three- and four-body nuclei
obtained from EFT potentials including 3NF. We agree with other methods at the 10 keV level.

Results for $S_{\cal M}(q,\omega)$ are often obtained by discretizing the continuum, where the Hamiltonian 
is represented on a finite  basis of square integrable functions and  is then diagonalized to  
obtain  eigenvalues $e_\nu$ and eigenfunctions  $|\nu\rangle$. In this way one achieves an ill defined 
discretized representation of $S_{\cal M}(q,\omega)$. On the contrary in the Lorentz integral
transform (LIT) approach~\cite{ELO94,EfL07} a continuum discretization can be properly used to reach the 
correct continuum spectrum (for various benchmark tests of the LIT approach we refer to Ref.~\cite{EfL07}).

In the LIT case one has
\begin{equation}\label{Gamma}
 {\cal L}_{\cal M}(q,\sigma,\Gamma)=  -\frac{1}{\pi}{\it Im}\langle 0| {\cal M}^\dagger (q)
 \frac{1}{\sigma-H+E_0+ i\Gamma}{\cal M} (q)|0\rangle\,, 
\end{equation}
where $\Gamma$ is finite (compare with Eq.~(\ref{S_M})). It is easy to prove that ${\cal L}_{\cal M}(q,\sigma,\Gamma)$ is connected to 
$S_{\cal M}(q,\omega)$ by an integral transform with a Lorentzian kernel
$K(\omega,\sigma,\Gamma)=\frac{\Gamma}{\pi}\frac{1}{(\omega+E_0-\sigma)^2+\Gamma^2}$,
\begin{equation}\label{trans}
{\cal L}_{\cal M}(q,\sigma,\Gamma)= \int d\omega\, K(\omega,\sigma,\Gamma) \,S_{\cal M}(q,\omega)\,.
\end{equation}
Since $\Gamma$ is finite the calculation of ${\cal L}_{\cal M}(q,\sigma,\Gamma)$ is 
a bound-state like problem and thus
it is legitimate to represent the Hamiltonian on a basis of square integrable functions, which
then leads to the following expression:
\begin{equation}
{\cal L}_{\cal M}(q,\sigma,\Gamma) =\frac{\Gamma}{\pi} \sum_{\nu=1}^N\frac{|\langle \nu|{\cal M} (q)|0\rangle|^2}
{(\sigma-e_\nu+E_0)^2+\Gamma^2}\,.
\end{equation} 
The number of basis functions $N$ depends in our EIHH calculation on
the maximal value $K_{\rm max}$ of the HH grand angular momentum quantum
number $K$. Note that the set ($e_\nu$, $|\nu\rangle$) is $\Gamma$-independent,
but that the convergence of ${\cal L}_{\cal M}$ is strongly correlated with
$\Gamma$: if $\Gamma$ is lowered a higher density of states is needed, hence
$K_{\rm max}$ and thus $N$ have to be increased. 
In our present case we reached convergence of ${\cal L}_{\cal M}$ 
with  $\Gamma$ as small as 5 MeV 
employing more than $10^5$ states $|\nu\rangle$.
Even if this is not sufficient to resolve the $^4$He* resonance width
of 270 keV, we are nevertheless able to determine the resonance energy $E_R$. In fact our discrete
spectrum shows as first excitation above the $^4$He ground state a very pronounced
state with strength $s_1(q)=|\langle 1|{\cal M}(q)|0\rangle|^2$, thus we identify the corresponding energy 
$e_1$ with $E_R$. We find the
following results: $E_R = -7.40(20)$ MeV (AV18+UIX) and $E_R=-7.50(30)$ MeV
(N$^3$LO+N$^2$LO). Note that error estimates are made by studying the EIHH convergence,
i.e. the $K_{\rm max}$ dependence of $E_R$ and that the $E_R$ value for N$^3$LO+N$^2$LO
is obtained extrapolating to higher $K_{\rm max}$ with an exponential ansatz $E(K_{\rm max})=E^{\infty}+a e^{-bK_{\rm max}}$ as in \cite{extrap}.
\begin{figure}[htb]
\centerline{\resizebox*{7.cm}{10.5cm}{\includegraphics[angle=0]{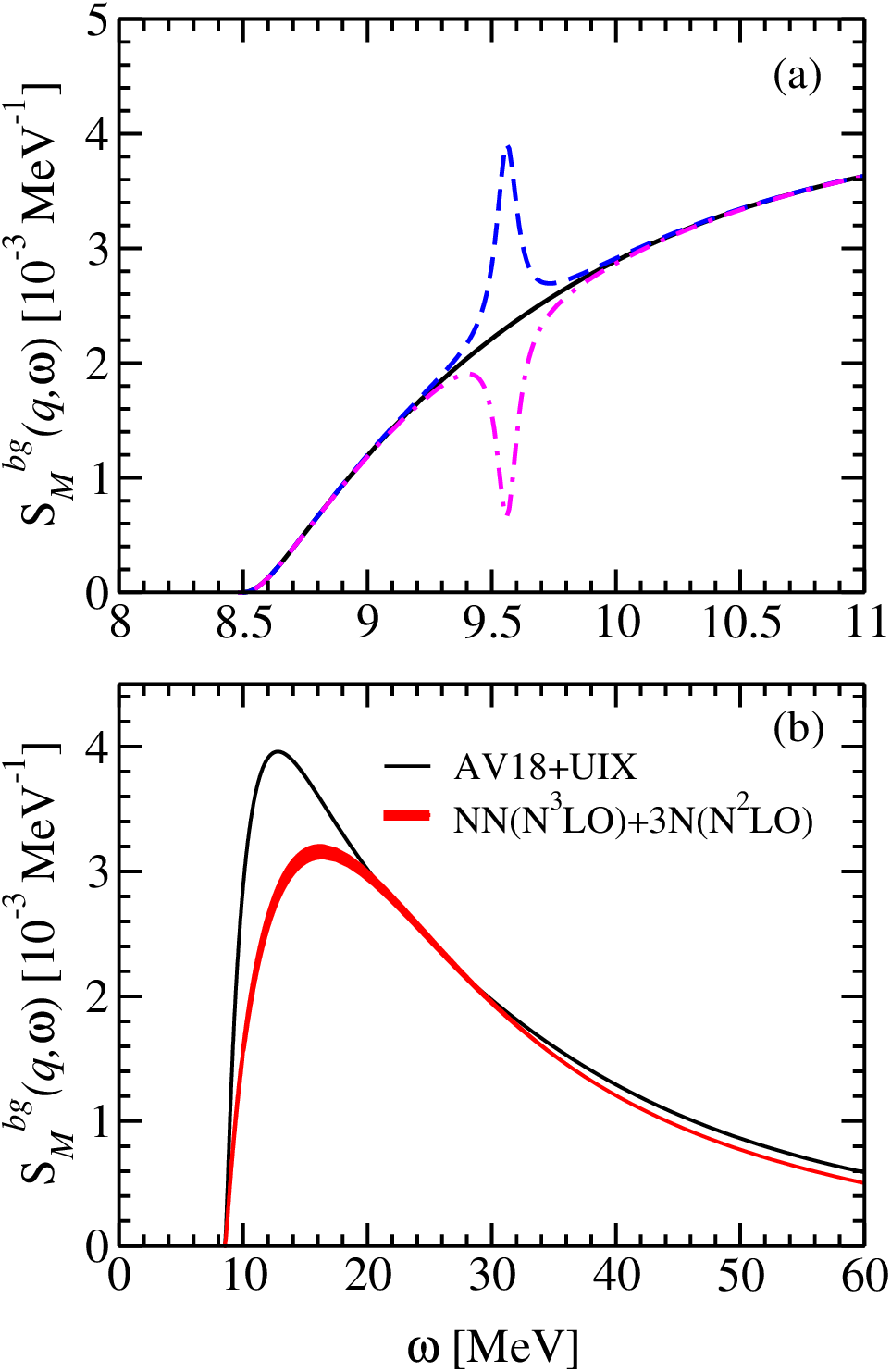}}}
\caption{(Color online) 
(a) ${\cal S}_{\cal M}^{bg}(q,\omega)$  at $q=1.5$ fm$^{-1}$ for AV18+UIX
obtained with different values of $f_R$ (see text): $f_R = |F_{\cal M}(q)|^2$ (full line),   
$f_R = 0.99 |F_{\cal M}(q)|^2$ (dashed line); $f_R = 1.01 |F_{\cal M}(q)|^2$ (dot-dashed line); 
(b) non-resonant background contribution ${\cal S}_{\cal M}^{bg}(q,\omega)$: 
AV18+UIX (full line); N$^2$LO+N$^3$LO (red band).
}
\label{fig_res_sub}
\end{figure}

In general one obtains the full strength $S(q,\omega)$ from the inversion of a converged LIT, but
one has to be aware that structures much smaller than $\Gamma$
cannot be resolved and thus a regularization procedure has to be used~\cite{TiA77,AnL05}. Our standard inversion method consists in an expansion of the response on a set 
of $I$ continuous functions and in fitting the calculated 
${\cal L}_{\cal M}(q,\sigma,\Gamma)$ on the corresponding linear 
combinations of the transformed basis functions~\cite{AnL05}. Note that 
the regularization consists in the fact that
$I$ should not become so large that structures
much smaller than $\Gamma$ appear in the inversion result. We implement many different basis and  choose the best fit  for a given $I$ (for example we use basis sets of the form $E^\beta exp(-\alpha E/i)$ with $i=1,...,I$ and different $\alpha$ values; $\beta$ is known from 
threshold behavior of the response, e.g. $\beta=1/2$ for $S_{\cal M}$).
For example, such a calculation has been made in Refs.~\cite{BaB09a,BaA07} for the full $^4$He longitudinal response beyond the
$^4$He* resonance.

In the presence of a narrow resonance, as in our case, an explicit resonance should be added to the basis,
e.g., a Lorentzian with free parameters $\gamma$ and $\omega_R$:
$[(\omega-\omega_R)^2+(\gamma/2)^2]^{-1}$.
If the LIT is determined with a sufficiently small $\Gamma$, then position, width, and strength of the resonance 
can be
determined in the inversion~\cite{Le08}. 
If we proceed in this way in our present case, imposing $\omega_R=E_R$, we obtain the 
best fits with $\gamma \rightarrow 0$. This reflects the absence of
states $|\nu\rangle$ in the vicinity of $E_R$. We can nonetheless determine,
besides $E_R$, also
the resonance strength $f_R(q)$. For this purpose we note that the above defined strength $s_1(q)$ is equal to 
the sum of $f_R(q)$ and a background contribution. Thus, formally, we can separate
the resonance contribution from ${\cal L}_{\cal M}$:
\begin{equation}
 {\cal L}^{bg}_{\cal M}(q,\sigma,\Gamma)={\cal L}_{\cal M}(q,\sigma,\Gamma)
              -\frac{\Gamma}{\pi}\frac{f_R(q)}{(\sigma-E_R+E_0)^2+\Gamma^2}\,.
\end{equation}

Now we proceed as follows. We assume a value for $f_R(q)$ and allow  a basis function
of Lorentzian shape centered at $E_R$ with $\gamma=100$ keV in the inversion. If the trial value for $f_R(q)$ is too small/large one finds
an inversion with a positive/negative resonant structure. The case where this vanishes corresponds
to the correct value of the transition form factor $|F_{\cal M}(q)|^2=f_R(q)/Z^2$ and the inversion result is just 
$S_{\cal M}^{\rm bg}(q,\omega)$ (see Fig.~\ref{fig_res_sub}).
We would like to emphasize that the results are almost $\gamma$-independent
 so long $\gamma$ remains small enough ($0 < \gamma \le 200$ keV) that
the Lorentzian approximates sufficiently well a $\delta$-function.  
For the AV18+UIX potential the relative size of the background reduction, about 8\%,
is roughly $q$ independent. For the N$^3$LO+N$^2$LO interactions
the reduction varies
between $13\%$ for $q= 0.25\,\rm{fm}^{-1}$ and $22\%$ for $q=2\,\rm{fm}^{-1}$. 

In Fig.~\ref{fig_res_sub}b, the  non-resonant monopole strength $S_{\cal M}^{\rm bg}$ is shown on
a larger energy range, into the far continuum region. One sees quite a difference between the results with the
EFT and the AV18+UIX forces. The former leads to a lower low-energy peak and tail
than the latter. These results show the power of the LIT approach,
which enables one to calculate the strength in the far four-body continuum by reducing a scattering-state problem
to a bound-state problem in a rigorous way.

\paragraph*{Analysis of the Results.}
The main findings of this Letter are the dramatic sensitivity of $F_{\mathcal M}(q)$ to the nuclear Hamiltonian
and the large deviations of realistic calculations from the available experimental data.
Even though one can contemplate the possibility of
systematic experimental errors, the fact that
the experimental results of Fig.~1 correspond to three different sets of data, makes it less likely.
Thus we will now list possible sources for theoretical uncertainties. 
\begin{table}
\caption{$|F_{\rm el}|$ and $s_1=|\langle 1|{\cal M} (q)|0\rangle|^2$ for $q=1.01$ fm$^{-1}$ 
as a function of the grandangular momentum $K_{\rm max}$ with
N$^3$LO+N$^2$LO~\cite{GaQ09}. 
\label{table_ff}}
\begin{center}
\begin{tabular}
{c c c cc} \hline\hline
$K_{\rm max}$ & 12 & 14 & 16 & 18 \\
\hline
$|F_{\rm el}|$                  & 0.6248 & 0.6244 &   0.6242 & 0.6241 \\
10$^{4}s_1/4\pi Z^2$      & 4.59   & 4.75   &   4.85   & 4.87 \\
\hline
\end{tabular}
\end{center}
\end{table}\\
 {\it (i)} Is our EIHH expansion sufficiently convergent? As shown in Table~\ref{table_ff} for a $q$-value of 1.01 fm$^{-1}$, we find 
an excellent convergence for both $F_{\rm el}$ and $s_1$.
\newline
{\it (ii)}  Are there relevant two-body corrections to the one-body operator of Eq.~(\ref{monopole})?
Such corrections are of relativistic order and appear also in EFT only
at 4th order \cite{Park} 
(also for $F_{\rm el}(q)$ such two-body terms are negligible below $q=2$ fm$^{-1}$~\cite{ViS07}).
\newline
{\it (iii)} Can additional 3NF terms change the picture? This is not excluded, however we notice that the 3NF effect at N$^2$LO on $F_{\cal M}(q)$ is rather mild (about 10\%).
\newline
{\it (iv)} Does the improper theoretical resonance position $E_R$ affect the $F_{\cal M}(q)$ result? 
Both our potential models (AV18+UIX, N$^3$LO+N$^2$LO) overestimate $E_R$
by almost the same amount (about $700$ keV), but still lead to quite different transition form factors. 
On the other hand, the simplified force model used by Hiyama {\it et al.}~\cite{HiG04}
reproduces the correct $E_R$ within $100$ keV, and also leads to a much better description of $F_{\cal M}(q)$.
One can envisage a correlation between the ability of a model
to reproduce $E_R$ and $F_{\cal M}$. In fact,
if one considers that $F_{\cal M}$ is the Fourier transform of
the transition density from $^4$He to $^4$He*, 
one can imagine that small differences in $E_R$ are reflected
in the resonant wave functions and yield   
larger differences in the transition density.
Similar conclusions have been drawn  
in Ref.~\cite{La09} in the study 
of $p\,$-$^3$H scattering. 
However, the resonant behavior of the nuclear scattering amplitude is barely visible
in the data, in contrast to
the electromagnetic probe that amplifies the resonance signal considerably (see Fig.~1 of Ref.~\cite{KoO83}).

\paragraph*{Conclusions.}
 We have calculated the isoscalar monopole $^4$He $\longrightarrow ^4$He* 
 transition form factor $F_{\mathcal M}(q)$ 
 with realistic nuclear forces (N$^3$LO+N$^2$LO, AV18+UIX).
 Unexpectedly the results are strongly dependent on the Hamiltonian. 
Therefore this observable is ideal for testing nuclear Hamiltonians.
As surprising as the large potential model dependence, is the fact that our $F_{\cal M}$ results are at 
variance with the experimental data, particularly large differences are found
in the case of the chiral forces.
It is very unlikely that  corrections to the isoscalar monopole operator
can lead to large effects. In order to clarify the situation it is highly desirable to have
a further experimental confirmation of the existing data and in particular
with increased precision.
On the theory side  
further insight could be gained by an analysis of sum rules, transition densities, effects of D-wave components and
different 3NFs. These issues will be the subject of future studies.

We would like to thank Thomas Walcher for his helpful discussions and
explanations about the experiments. We would like to thank Michele Viviani for providing
us with his theoretical results for $F_{\rm el}$. 
Acknowledgments of financial support are given to
Natural Sciences and Engineering Research Council (NSERC) and
the National Research Council of Canada, S.B.,
the Israel Science Foundation (Grant number 954/09), N.B., 
the MIUR grant PRIN-2009TWL3MX, W.L. and G.O..  
We would also like to thank the INT for its hospitality during the preparation of
this work (INT-PUB-12-050).

\end{document}